\documentclass[aps]{revtex4}
\usepackage{graphicx}
\usepackage{setspace}
\begin{document}
\begin{spacing}{2.0}

\title{Universal spin-glass behaviour in bulk LaNiO$_{2}$, PrNiO$_{2}$ and NdNiO$_{2}$}
\author{Hai Lin$^*$, $^1$ Dariusz Jakub Gawryluk, $^2$ Yannick Maximilian Klein, $^2$ Shangxiong Huangfu, $^1$ Ekaterina Pomjakushina, $^2$ Fabian von Rohr, $^3$ and Andreas Schilling $^1$}
\affiliation{$^1$ Department of Physics, University of Z$\ddot{u}$rich, Winterthurerstrasse 190, CH-8057 Z$\ddot{u}$rich, Switzerland\\
$^2$ Laboratory for Multiscale Materials Experiments (LMX), Paul Scherrer Institute (PSI), Forshungstrasse 111, CH-5232 Villigen, Switzerland\\
$^3$ Department of Chemistry, University of Z$\ddot{u}$rich, Winterthurerstrasse 190, CH-8057 Z$\ddot{u}$rich, Switzerland}
\email{hailin@physik.uzh.ch}

\date{\today}

\begin{abstract}
Motivated by the recent discovery of superconductivity in infinite-layer nickelate thin films, we report on a synthesis and magnetization study on bulk samples of the parent compounds ${R}$NiO$_{2}$ (${R}$=La, Pr, Nd). The frequency-dependent peaks of the AC magnetic susceptibility, along with remarkable memory effects, characterize spin-glass states. Furthermore, various phenomenological parameters via different spin glass models show strong similarity within these three compounds as well as with other rare-earth metal nickelates. The universal spin-glass behaviour distinguishes the nickelates from the parent compound CaCuO$_{2}$ of cuprate superconductors, which has the same crystal structure and $d^9$ electronic configuration but undergoes a long-range antiferromagnetic order. Our investigations may indicate a distinctly different nature of magnetism and superconductivity in the bulk nickelates than in the cuprates. 
\end{abstract}\maketitle

\section{Introduction}
The nickelates are widely believed to be promising candidates for unconventional superconductors because of their very similar structures and possibly related electronic states to the cuprates. Superconductivity was realized in nickelate thin films. Nd$_{1-x}$Sr$_{x}$NiO$_{2}$ films grown on SrTiO$_{3}$ were found to be superconducting with $T_c$ = 9 $\sim$ 15 K\cite{NdSrNiO2nature}, and corresponding Pr$_{1-x}$Sr$_{x}$NiO$_{2}$ films with $T_c$ = 7 $\sim$ 12 K\cite{PrSrNiO2}. Very recently, Sr- and Ca-doped LaNiO$_2$ thin films\cite{LaSrNiO2SC, LaCaNiO2SC} have also been shown to be superconducting.This discoveries have aroused intensive research and discussions, but several open questions remain. 

First, superconductivity is absent in bulk samples, e.g. in bulk Nd$_{1-x}$Sr$_{x}$NiO$_{2}$\cite{bulk1, bulk2} and Sm$_{1-x}$Sr$_{x}$NiO$_{2}$\cite{SmSrNiO2}, and no signs of long-range magnetic orders have been observed. This discrepancy between non-superconducting bulk samples and superconducting thin films has not been well understood so far. Off-stoichiometry defects or chemical inhomogeneity in bulk samples are likely suspects to hamper superconductivity, while the role of the SrTiO$_3$ substrate has also been considered\cite{interface, thickness, surface1, surface2}. Strain effects may play an important role in the superconductivity of thin films, where a 3$\%$ larger $c$-axis lattice constant affects the Ni $d_{z^2}$ orbital which may be crucial for nickelate superconductivity\cite{Pickett, strain, thickness}. 
Second, the parent phase $R$NiO$_{2}$ shows several differences from the isostructual parent compound CaCuO$_{2}$, which can be a high-temperature superconductor upon doping\cite{SrLaCuO2, CaSrCuO2}. Although NdNiO$_{2}$ and CaCuO$_{2}$ have the same infinite-layer structure and the same 3${d}^9$ configuration, NdNiO$_{2}$ is a conductor\cite{NdNiO2} (or a weak insulator\cite{NdSrNiO2nature,NdNiO2,bulk1,bulk2,dome1,dome2}) without any long-range magnetic order, whereas CaCuO$_{2}$ is an antiferromagnetic (AFM) Mott insulator\cite{CaCuO2}. The superconductivity in the nickelate may therefore be far away from the well-known scenario in high-temperature superconductors, where AFM spin fluctuations have been suggested to act as the medium to promote superconductivity. More distinctions, such as a larger charge transfer gap and smaller AFM exchange energy have been found in experiments\cite{XPSNdNiO2, XAS} and theoretical calculations\cite{NiisnotCu, SimilaritiesDifferences, Norman, criticalnature, substantialhybridization}. 
Moreover, the role of the 4$f$ electrons of the $R$ ions in $R$NiO$_{2}$ is still unknown. Theoretical calculations propose that the 4$f$ electrons have an important influence on the electronic structure through 4$f$-5$d$ interactions\cite{roleof4f, nonzerofness, felectron}, although the doped LaNiO$_2$ thin films have recently also been shown to superconduct\cite{LaSrNiO2SC, LaCaNiO2SC}.

A good starting point to clarify these questions would be to study the magnetism of the parent phases. The difficulty we face is to obtain bulk samples of $R$NiO$_{2}$, especially PrNiO$_2$, and to distinguish the intrinsic magnetism therein. In this work, we synthesized bulk LaNiO$_{2}$, PrNiO$_{2}$ and NdNiO$_{2}$ samples and systematically investigated their magnetic properties. We show that magnetic memory effects and frequency-dependent spin freezing peaks in AC magnetization reveal a universal spin-glass behaviour in $R$NiO$_{2}$.

\section{Method}
The synthesis of ${R}$NiO$_{2}$ (${R}$=La, Pr, Nd) samples was conducted in two stages, according to the same procedure as in previous reports\cite{LaNiO2, NdNiO2, bulk1, bulk2, SmSrNiO2}. Here, we put an emphasis on bulk PrNiO$_{2}$, which is obtained for the first time up to our best knowledge. In the first stage, high-quality PrNiO$_{3}$ samples were grown under a high-pressure oxygen atmosphere\cite{PrNiO3}. In the second stage, the precursor PrNiO$_{3}$ samples were topochemically reduced by applying the following route, mixed with CaH$_2$, sealed in quartz tubes and heated at 270 $^{\circ}$C for 40 hours. Afterwards, the mixtures were washed by NH$_4$Cl in ethanol to remove the extra CaH$_{2}$ and Ca(OH)$_{2}$. The resulting black powders were PrNiO$_{2}$. LaNiO$_{2}$ and NdNiO$_{2}$ were synthesized in the same way. The pictures taken by a scanning electron microscope(SEM) indicate that all these samples consist of 1 $\sim$ 5 ${\rm\mu}$m sized particles, as shown in the insets in Figure~\ref{fig1}. Last but not least, the samples were stored in sealed centrifuge tubes and handled in the N$_2$ atmosphere in order to avoid degradation caused by water or oxygen in the air. 

We made an elemental analysis with both energy-diffraction spectra (EDX) and thermal gravity(TG). The EDX measurements as well as the SEM pictures were performed on a Zeiss GeminiSEM 450 with a beam energy of 10 keV. The obtained atomic ratios were $R$ : Ni = 1.01/1.04/1.02 for La/Pr/NdNiO$_{2}$, respectively. The TG analysis was done to determine the oxygen content in the respective oxides. Considering that the ${R}$NiO$_{2}$ samples would be totally reduced to ${R_2}$O$_{3}$ and Ni by H$_{2}$ in 700 $^{\circ}$C, and by comparing precisely the mass before and after the reduction, the oxygen contents were deduced to 2.05/1.95/1.99$\pm$0.04 for La/Pr/NdNiO$_{2}$, respectively. 
Altogether, the chemical composition of ${R}$NiO$_{2}$ we obtained is very close to the nominal compounds $R$NiO$_{2}$ within experimental resolution.

The X-ray diffraction (XRD) measurements were performed on a Bruker AXS D8 Advance diffractometer (Bruker AXS GmbH, Karlsruhe, Germany), equipped with a Ni-filtered Cu K$\alpha$ radiation and a 1D LynxEye PSD detector. The Rietveld refinement analysis\cite{Rietveld} of the diffraction patterns was performed with the package FULLPROF SUITE (version March-2019). All the magnetic measurements including the direct current (DC), alternating current (AC) magnetic susceptibility and memory effect, were performed in a Magnetic Properties Measurement System (Quantum Design MPMS-3).

\section{Results}
\begin{figure}
  \includegraphics[width=8.5cm]{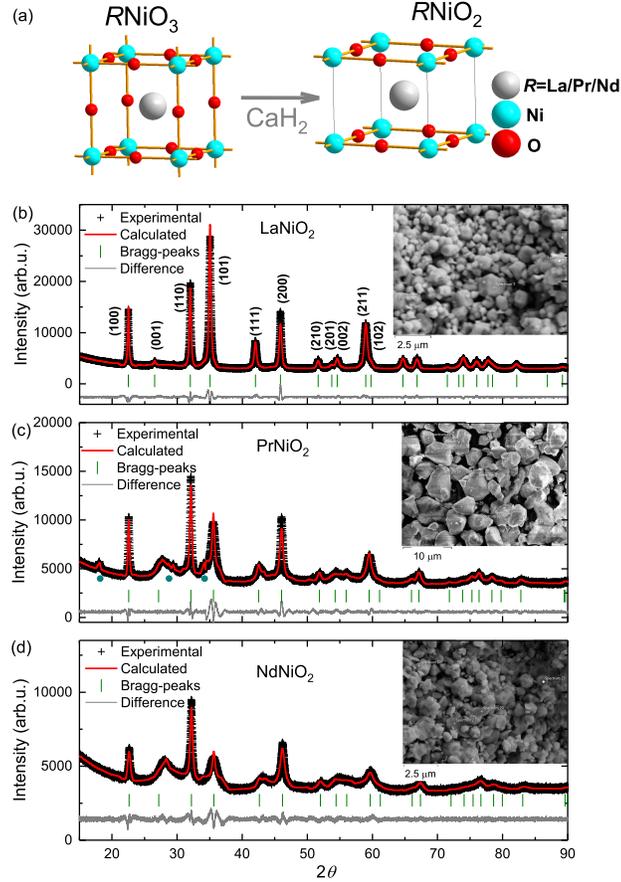}
\caption{Schematic structures of (a) ${R}$NiO$_{3}$ and ${R}$NiO$_{2}$ (${R}$ = La, Pr, Nd). Experimental (black crosses) and calculated (red lines, via Rietveld method) powder X-ray diffraction patterns of the (b) LaNiO$_{2}$, (c) PrNiO$_{2}$ and (d) NdNiO$_{2}$ polycrystalline samples are shown, together with the Bragg peak positions (vertical green lines). The green dots represent the peaks from residual Ca(OH)$_2$. The insets display the respective SEM pictures. } \label{fig1}
\end{figure}

We performed XRD measurements to characterize the crystal structures of the as-synthesized $R$NiO$_{2}$ samples. The corresponding data are shown in Figure~\ref{fig1}, together with refinements via the Rietveld method and the calculated Bragg peaks. As illustrated in Figure~\ref{fig1}(a), after extracting the apical oxygen in $R$NiO$_{3}$, the NiO$_2$-planes in $R$NiO$_{2}$ are separated by rare-earth metal ions layer by layer. The reduction might in principle be incomplete or introduce crystalline disorder or even destroy the NiO$_2$-planes into elemental Ni. However, the calculated curves fit well with the experimental ones, and all $R$NiO$_{2}$ compounds can be successfully refined as tetragonal phases with the same space group $P4/mmm$. These XRD results indicate that no significant distortions were introduced in the NiO$_2$-plane by the reduction, and the compounds can indeed be considered as the infinite-layer phase. The crystal parameters obtained from the Rietveld refinements are presented in Table1. The $c$-axis parameters of LaNiO$_{2}$, PrNiO$_{2}$ and NdNiO$_{2}$ are 3.363 ${\rm \mathring{A}}$, 3.285 ${\rm \mathring{A}}$ and 3.279 ${\rm \mathring{A}}$, respectively. They are close to the reported values for bulk samples\cite{LaNiO2, LaNiO2-2, LaNiO2-3, NdNiO2, bulk1}, but 3$\%$ smaller than in thin films\cite{NdSrNiO2nature, PrSrNiO2}. The different $c$ values in our samples can be simply attributed to the different radii of rare-earth metal ion for coordination number $N$ = 8 that $r$(La$^{3+}$) = 1.15 ${\rm \mathring{A}}$, $r$(Pr$^{3+}$) = 1.11 ${\rm \mathring{A}}$ and $r$(Nd$^{3+}$) = 1.10 ${\rm \mathring{A}}$\cite{ionicradii}. 

To illustrate the anisotropic widths of the Bragg reflections in $R$NiO$_{2}$, the three main reflection classes $\langle{100}\rangle$, $\langle{110}\rangle$, $\langle{101}\rangle$ are chosen and the corresponding FWHM (full-widths at half-maximum) of the corresponding X-ray diffraction peaks are presented in Table1. It can be seen that the sharper reflections are those from diffraction planes $\langle{hk0}\rangle$, while the broad peaks belong to the class where $l$ has a non-zero value. In other words, the coherent domains along the [001] direction are smaller than along [100]. This anisotropic shrinking of the crystalline size after topotactic reduction has already been observed in previous studies on nickelates\cite{LaNiO2, LaNiO2-2, bulk1, bulk2, NdNiO2, Nd4Ni3O8} and cobaltates\cite{LaSrCoO}, and are characteristic for systems with anisotropic particle size, stacking disorder, or uniaxial strain in layered compounds. In the $R$NiO$_2$ system, this anisotropy may be due to the stronger intra-layer than inter-layer bonding, making the structure to more likely break up along the [001] direction. On the other hand, the crystalline sizes become smaller from LaNiO$_{2}$, PrNiO$_{2}$ to NdNiO$_{2}$, especially for $\langle{101}\rangle$. This indicates an increasing instability of the $R$NiO$_2$ structure from $R$=La, Pr to Nd, which could be caused by the smaller lattice parameters and consequently the stronger inter-layer Ni-Ni interactions. This would be consistent with previous research on Nd$_{1-x}$Sr$_x$NiO$_2$ where a higher Sr-doping level with a larger $c$-axis parameter exhibits a sharper and more intense $\langle{101}\rangle$ peak\cite{bulk1, bulk2}. 

It is worth noting that there are no observable nickel impurities in these samples. The nickel was one of the prime suspects responsible for the FM-like transition at low temperature as well as the weak magnetic hysteresis in bulk Nd$_{0.8}$Sr$_{0.2}$NiO$_{2}$ and Sm$_{0.8}$Sr$_{0.2}$NiO$_{2}$\cite{bulk1, bulk2, SmSrNiO2}. To further exclude the possibility of nickel impurities, we re-oxidized the samples back to $R$NiO$_{3}$ and did not observe any NiO. In addition, we measured the nickel powders synthesized using the same method and did not find any frequency-dependent behaviour as we describe below.  Combining all of this evidence, we may claim to have no observable amount of crystalline nickel impurities in the as-synthesized LaNiO$_{2}$, PrNiO$_{2}$ and NdNiO$_{2}$ samples, and the magnetic measurements to be presented below reveal intrinsic properties of these poly-crystalline samples. The broad feature around 2$\theta$ = 28$^\circ$ for PrNiO$_{2}$ and NdNiO$_{2}$ of unknown origin which also occurs in the corresponding data of other works\cite{bulk1, NdNiO2, XPSNdNiO2} is not likely to affect this result, as the LaNiO$_2$ without this feature still show a similar magnetic behaviour.

\begin{figure}
  \includegraphics[width=8.5cm]{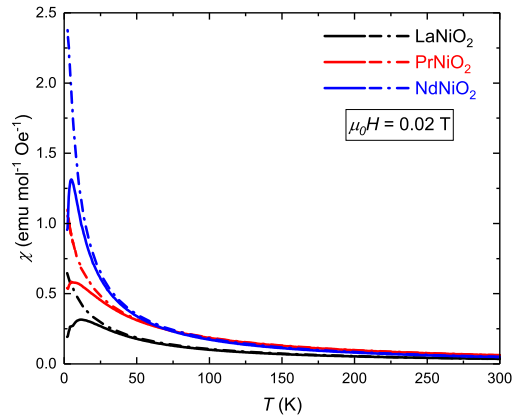}
\caption{DC magnetic susceptibilities at 0.02 T for LaNiO$_{2}$ (black), PrNiO$_{2}$ (red) and NdNiO$_{2}$ (blue), respectively. The solid lines are measured for zero-field cooling (ZFC) and the dashed ones for field cooling (FC) processes.} \label{figxT}
\end{figure}
\begin{figure}
  \includegraphics[width=8.5cm]{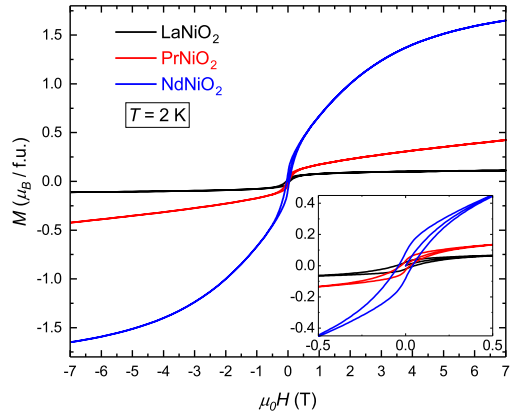}
\caption{Magnetic-hysteresis loops (MHLs) at 2 K for ${R}$NiO$_{2}$ (${R}$ = La, Pr, Nd). The inset is the enlarged view for low magnetic fields.} \label{figMH}
\end{figure}
The DC magnetic susceptibilities $\chi$($T$) under a magnetic field of 0.02 T for the ${R}$NiO$_{2}$ samples (${R}$ = La, Pr, Nd) are shown in Figure~\ref{figxT}. It is obvious that there are peak-like structures at 11.8 K for LaNiO$_{2}$, 6.7 K for PrNiO$_{2}$ and 4.8 K for NdNiO$_{2}$, respectively. Taking into account of the huge difference between ZFC and FC data above the peak temperature, we interpret this peak-like temperature as a spin-glass freezing process, and the corresponding peak temperature is in fact a spin freezing temperature $T_f$. The weak hysteresis shown in Figure~\ref{figMH} in bulk ${R}$NiO$_{2}$ samples can also be considered as a feature of a spin-glass state.

\begin{figure}
  \includegraphics[width=8.5cm]{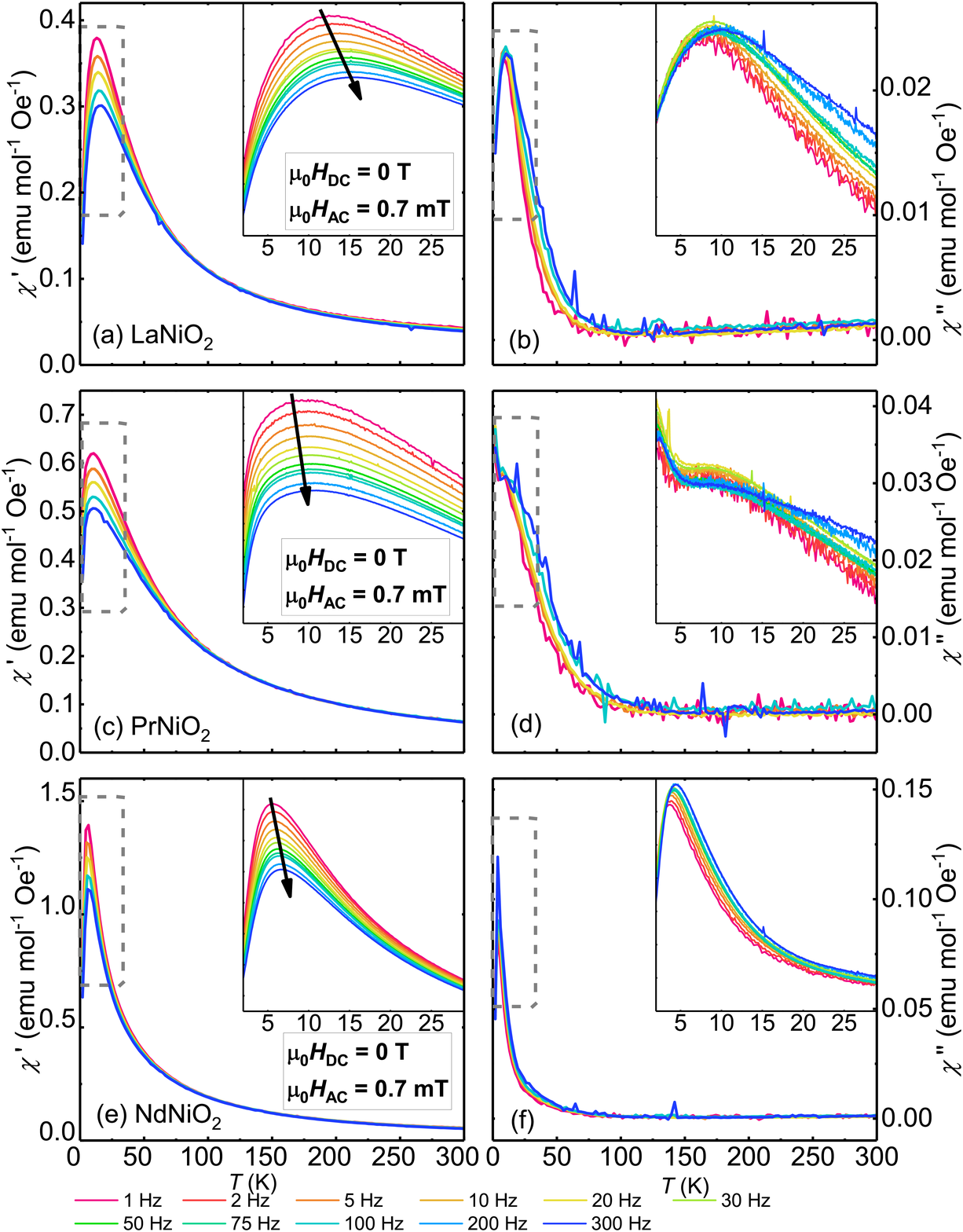}
\caption{Real part $\chi'$ and imaginary part $\chi''$ of the AC magnetic susceptibility of ${R}$NiO$_{2}$ (${R}$ = La, Pr, Nd) measured at different frequencies with zero external DC magnetic field. The insets are enlarged drawings from 2 K to 30 K. The black arrows illustrate the shifts of the $\chi'$ peak temperatures $T_p$.} \label{figAC}
\end{figure}
The AC susceptibility technique can be used as a probe of magnetic dynamics to distinguish spin-glass systems from other, long-range ordered systems\cite{AC}. The results of systematic AC susceptibility measurements on ${R}$NiO$_{2}$ (${R}$ = La, Pr, Nd) samples for AC frequencies ranging from 1 Hz to 300 Hz are plotted in Figure~\ref{figAC}. The applied DC magnetic field is 0, and the AC magnetic fields have the same amplitude of 0.7 mT. In all cases, we observe distinct peaks in the real parts $\chi'$ of the magnetic susceptibility, and an upward shift of the peak temperatures with increasing frequency $f$. At the same time, the imaginary parts $\chi''$ indicate strong dissipation starting already below $\sim$100 K, with a maximum around the peak temperatures of the respective $\chi'$ data, with the exception of PrNiO$_{2}$ where the peak structure in $\chi''$ is less clear. Such a behavior is characteristic of spin-glass systems and distinguishes them from disordered AFM systems, in which $\chi''$ remains zero even below the transition temperature\cite{UCu4Pd,URh2Ge2,AuMn}. 
Apart from these, however, there is an intriguing deviation from the typical spin-glass behaviour: the frequency-dependence disappears only at temperatures as high as 70 K$\sim$100 K, nearly 10 times higher than the peak temperature. This may indicate a complex magnetism in $R$NiO$_2$, similar to that we have reported for the square planar nickelate Pr${_4}$Ni$_3$O$_{8}$\cite{Pr438}. Multiple magnetic disorders may participate in the frustration of magnetic interactions, one that is dominating the freezing process at $T_p$, accompanied by another, more gradual one at a higher temperature. Interestingly, this temperature region is close to the metal-insulator transition temperature observed in the electric resistivities of ${R}$NiO$_{2}$ thin films\cite{LaNiOresistivity, PrSrNiO2, NdSrNiO2nature}. This may imply an underlying relationship between the magnetic disorder and insulating behaviour.   

\begin{figure}
  \includegraphics[width=8.5cm]{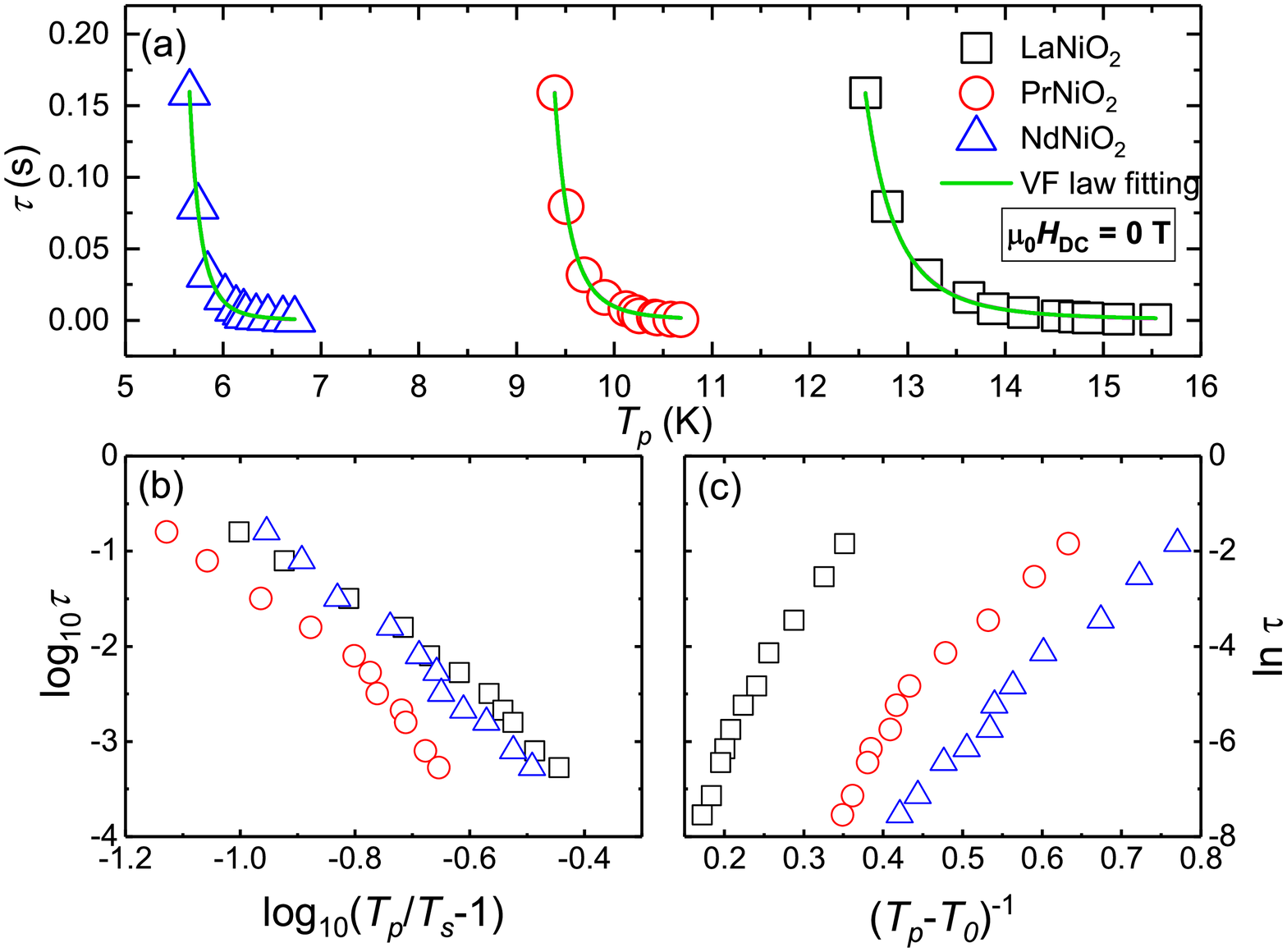}
\caption{(a) Measured oscillation periods $\tau = 2 \pi f^{-1}$ as functions of peak temperatures $T_p$ of the AC $\chi'$ curves for LaNiO$_{2}$ (black squares), PrNiO$_{2}$ (red circles) and NdNiO$_{2}$ (blue triangles). The green lines present the fitting results by VF laws. The scaling plots, corresponding to CDS theory and VF laws are shown in (b) as log$_{10}\tau$ $\sim$ log$_{10}$($T_p$/$T_s$-1) and (c) as ln$\tau\sim(T_p-T_0)^{-1}$, respectively.} \label{figTp}
\end{figure}

The relative variation of the freezing temperature $T_f$ per frequency decade $\delta = \Delta T_f / (T_f \Delta$ log$_{10}f$), is often employed as a parameter to compare different kinds of spin-glass systems. It reflects the strength of magnetic interactions between the underlying entities. Here, the peak temperature $T_p$ is identified as the spin glass freezing temperature $T_f$. The values of $\delta$ obtained in this way are all very close, namely 0.077, 0.049, and 0.064 for LaNiO$_{2}$, PrNiO$_{2}$ and NdNiO$_{2}$, respectively. These values are intermediate between canonical spin-glass systems such as CuMn\cite{CuMn} ($\delta$ = 0.005) and superparamagnetic systems such as $\alpha$-[Ho$_2$O$_3$(B$_2$O$_3$)]\cite{Mydosh} ($\delta$ = 0.28). They fall within the order of magnitude usually observed for cluster spin glasses, such as Ni-doped La$_{1.85}$Sr$_{0.15}$CuO$_4$\cite{LaSrCuNiO} ($\delta$ = 0.012), PrRhSn$_3$\cite{PrRhSn} ($\delta$ = 0.086), and therefore $R$NiO$_2$ resembles the case of cluster spin glasses. 

To obtain a more detailed insight in the spin-glass behaviour, we have studied the relation between $T_p$ and the fluctuation time $\tau=1/2{\pi}f$, as presented in Figure~\ref{figTp}(a). The critical dynamical scaling (CDS) theory predicts a power law: 
\begin{equation}
\tau = \tau_* (\frac{T_p}{T_s}-1)^{-z\nu} .
\label{equation1}
\end{equation}
In this equation, $\tau$ describes the fluctuation time scale, $z$ the dynamic critical exponent reflecting the correlated dynamics diverging as $\tau\propto\xi^z$, and $\nu$ the static critical exponent describing the divergence of the correlation length $\xi\propto(T-T_s)^{-\nu}$. The $\tau_*$ is the relaxation time of a fluctuating entity flip, and $T_s$ $\textless$ $T_p$ is the static freezing temperature as $f$ tends to zero. From Table \ref{tbl}, the resulting $\tau_*$ of $R$NiO$_2$ is 10$^{-5}\sim$10$^{-6}$ s, far larger than the typical values for a canonical spin glass, 10$^{-12}\sim$10$^{-14}$ s\cite{Mydosh}, and still larger than the values of the reported cluster spin-glasses 10$^{-7}\sim$10$^{-10}$ s\cite{Mydosh, LaSrCuNiO}. Such a high value of $\tau_*$ implies a slow dynamics of the fluctuating entities of $R$NiO$_2$. The resulting $z\nu$ of $R$NiO$_2$ lies around 4$\sim$5, at the lower bound found in typical spin glass systems 4$\sim$12\cite{Mydosh}.   

Alternatively, the spin glass state can also be described by the VF law. By introducing an empirical ''true transition point or anomaly'' at $T_0$ into the conventional Arrhenius law, the Vogel-Fulcher (VF) law can be written as 
\begin{equation}
\tau = \tau_0  {\rm exp}(\frac{E_a/k_B}{T_p-T_0}) .
\label{equation2}
\end{equation}
In this equation, $\tau_0$ is essentially equivalent to $\tau_*$ (see Table \ref{tbl}), and $E_a$ is an energy barrier. Considering $E_a$ in general to be correlated with the transition temperature for different spin-glass systems, $E_a$/$k_BT_0$ is a more meaningful measure of the coupling between the interacting entities. An $E_a$/$k_BT_0\gg$ 1 would indicate a weak coupling, and $E_a$/$k_BT_0\ll$1 a corresponding strong coupling. The value we obtain for $R$NiO$_2$ is 2$\sim$3, in the intermediate regime. The scaling plots according to the CDS theory and the VF law are shown in Figures~\ref{figTp} (b) and (c). These scatter plots should ideally be linear, and they imply a fair fitting to both models. All the parameters obtained from this procedure are shown in Table \ref{tbl}.

\begin{table}
  \caption{Systematic comparison between $R$NiO$_2$ ($R$=La, Pr, Nd) and Pr$_4$Ni$_3$O$_8$\cite{Pr438}}
  \label{tbl}
  \begin{tabular}{lllllll}
    \hline
    \hline
    compound & LaNiO$_2$ & PrNiO$_2$ & NdNiO$_2$ & Pr$_4$Ni$_3$O$_8$ \\
    \hline
    $R^{3+}$ configuration & 4$f^0$ & 4$f^2$ & 4$f^3$\\
    $a$ at 300 K (${\rm \mathring{A}}$) & 3.9582(1) & 3.9403(3) & 3.9279(4)\\
    $c$ at 300 K (${\rm \mathring{A}}$) & 3.3632(3) & 3.2845(8) & 3.2793(11)\\
    $a$/$c$ at 300 K & 1.18 & 1.20 & 1.20  \\
    $R_p$ ($\%$) & 2.1 & 2.1 & 1.9 \\
    $R_{wp}$ ($\%$) & 3.0 & 3.4 & 2.6 \\
    \hline
    FWHM$\langle{100}\rangle$ ($^{\circ}$) & 0.21 & 0.23 & 0.29\\
    FWHM$\langle{110}\rangle$ ($^{\circ}$) & 0.28 & 0.31 & 0.38\\
    FWHM$\langle{101}\rangle$ ($^{\circ}$) & 0.41 & 0.87 & 1.03\\
    \hline
    $T_p$ at 0.02 T (K) & 11.8 & 6.7 & 4.8\\ 
    $\delta$ & 0.077 & 0.049 & 0.064 & 0.057\\
    $\tau_*$ (s) & 2.8$\times$10$^{-5}$ & 2.7$\times$10$^{-6}$ & 1.9$\times$10$^{-6}$ & $\sim$ 10$^{-6}$\\
    $T_s$ (K) & 11.4 & 8.7 & 5.1 & 68.3\\
    $z\nu$ & 3.7 & 4.2 & 5.2 & 3.8\\
    $\tau_0$ (s) & 1.7$\times$10$^{-5}$ & 6.7$\times$10$^{-6}$ & 1.3$\times$10$^{-6}$ & $\sim$ 10$^{-6}$\\
    $T_0$ (K) & 9.7 & 7.8 & 4.4 & 60.1\\
    $E_a$/$k_B$ (K) & 26.1 & 15.9 & 15.2 & 130\\
    \hline
    $\tau_r$ (s) & 1.4$\times$10$^{3}$ & 1.4$\times$10$^{3}$ & 1.6$\times$10$^{3}$ & 1.8$\times$10$^{3}$\\
    $\beta$ & 0.49 & 0.47 & 0.45 & 0.49\\
    \hline
    \hline
  \end{tabular}
\end{table}

\begin{figure}
  \includegraphics[width=8.5cm]{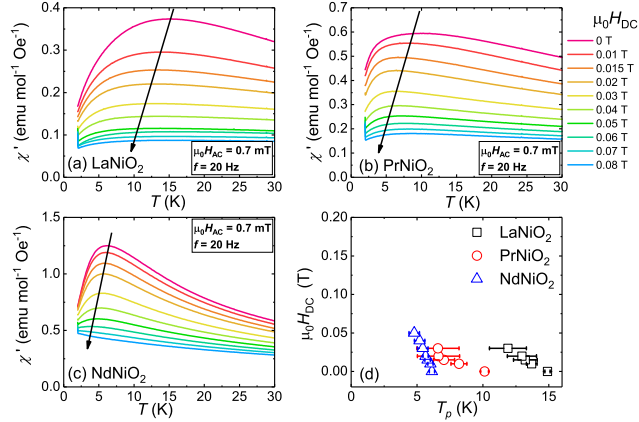}
\caption{Real part $\chi'$($T$) of the AC magnetic susceptibility of (a) LaNiO$_{2}$, (b) PrNiO$_{2}$ and (c) NdNiO$_{2}$ measured in different external DC magnetic fields from 0 to 0.08 T. (d) Peak temperature $T_p$ of ${R}$NiO$_{2}$ as a function of $H$. Due to the magnetic field broadening of the peaks in $\chi'$($T$), the $T_p$ for $\mu_0H_{DC}$ $\textgreater$ 0.05 T cannot be determined very accurately (see error bars).}
\label{figHTp}
\end{figure}

Furthermore, the DC magnetic field dependence of AC susceptibility $\chi'$ around the freezing temperature is shown in Figure~\ref{figHTp} for (a) LaNiO$_{2}$, (b) PrNiO$_{2}$ and (c) NdNiO$_{2}$, respectively. The applied DC magnetic fields range from 0 to 0.08 T. The amplitude of AC magnetic field is 0.7 mT and the frequency is 20 Hz. One can see that the peaks in $\chi'$ are significantly suppressed with increasing DC magnetic field and may even vanish in moderately high DC magnetic fields of the order of 1 T or even less. The corresponding peak temperatures $T_p$, and along with them most likely also the spin freezing temperatures $T_f$ $\textless$ $T_p$, decrease rapidly as it is usually observed in spin glasses (see Figure~\ref{figHTp} (d)) \cite{PrRhSn, NdSrMnO3, LuFe2O4, EuRu1222, CrFeGa}. This may explain why a recent NMR study \cite{NMRLaNiO2} performed in $\mu_0H$ = 12 T did not reveal any signs for spin-freezing in LaNiO$_2$ down to $T$ = 0.24 K. 

\begin{figure}
  \includegraphics[width=8.5cm]{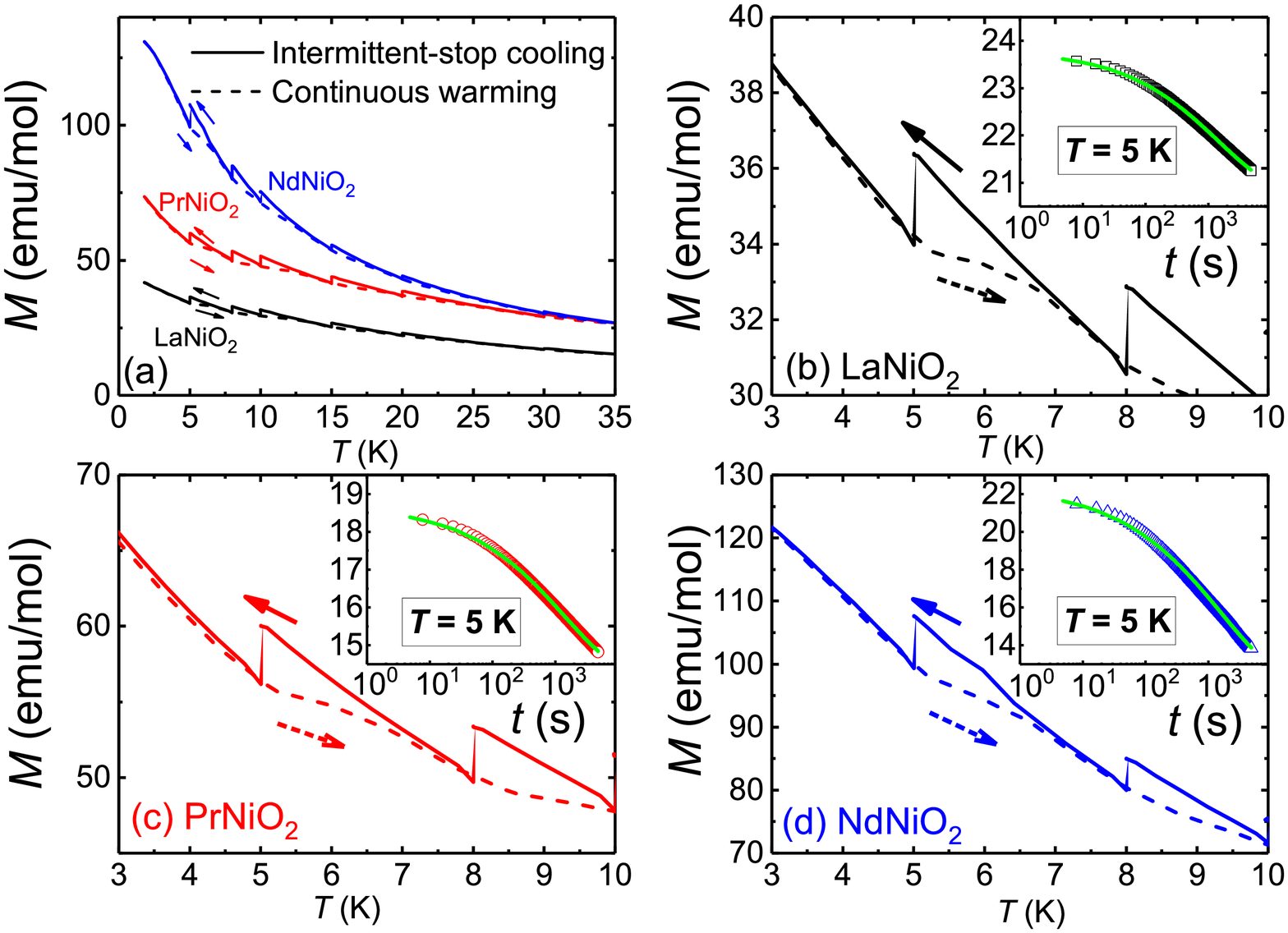}
\caption{(a) Memory effects in ${R}$NiO$_{2}$ (${R}$ = La, Pr, Nd). The solid lines are measured in the intermittent-stop cooling process and the dashed ones in the continuous warming process. The enlarged views are respectively shown in (b), (c), (d). The insets illustrate the time evolution of magnetic susceptibility measured at 5 K, immediately after turning off the external magnetic field. And the green lines represent the fitting results by stretched exponential functions.} \label{figmemory}
\end{figure}

A solid evidence for the spin glass behaviour in ${R}$NiO$_{2}$ is the memory effect in the magnetization. Memory-effect measurements were performed using two different protocols, an intermittent stop cooling process (ISC-process) and a continuous warming process (CW-process). In the first one, the sample is cooled in a magnetic field of 50 Oe from 300 K down to 1.8 K. Some intermittent stops are applied at certain temperatures. During these stops, the external field is switched off quickly and the temperature is kept constant for 5000 seconds while taking magnetization data simultaneously. Afterwards, the field is switched back to the original value and the sample further cooled down. Eventually at 1.8 K, this process is immediately followed by warming continuously from 1.8 K to 300 K. 
A typical feature of disordered spin glasses is a relatively long timescale to reach an equilibrium state when the magnetic field is changed. The magnetization in the glassy region would show decays in the ISC-process, and peculiar history dependent effects in the CW-process. As shown in Figure~\ref{figmemory}, we can see a significant memory effect in all ${R}$NiO$_{2}$ compounds, in which the CW-process data (dashed lines) show distinct features at the previous stopping temperatures, and therefore reflect the history of the previous ISC-process.
Another typical feature of spin glasses is that the magnetization below $T_p$ decays slowly after the magnetic field is turned off. The insets of Figure~\ref{figmemory} (b), (c), (d) show the magnetic moment measured at 5 K for 5000 seconds after setting the field to zero. The time decay of the magnetization can be fitted perfectly by a stretched exponential function,
\begin{equation}
M(t) = M_0+M_g {\rm exp}[-(\frac{t}{\tau_r})^\beta] ,
\label{equation3}
\end{equation}
where $M_0$ is an intrinsic magnetization, $M_g$ is related to a glassy component of magnetization, $\tau_r$ is the characteristic relaxation time constant, and $\beta$ is the stretching exponent, which has values between 0 and 1 and is a function of temperature only. The fitting parameters $\tau_r$ and $\beta$ for ${R}$NiO$_{2}$ and Pr$_4$Ni$_3$O$_8$ are presented in Table \ref{tbl}, and the respective fitting results are shown as the green lines in Figure~\ref{figmemory}. All these phenomena above are strongly indicative of a typical spin glass behaviour in ${R}$NiO$_{2}$. 

It is necessary to verify that the observed spin-glass behavior in these nickelate samples is not caused by nickel impurities. On the one hand, the structural and elemental analysis showed no crystalline nickel in the samples. On the other hand, after a complete reduction to $R_2$O$_3$ and Ni as we mentioned above for the thermal gravity experiments, the resulting products show a complete absence of spin-glass behavior (see Supplementary Material). Even if there had been elemental nickel impurities (crystalline or amorphous) induced by the initial reduction process and mimicking a spin-glass sate in ${R}$NiO$_{2}$, the same feature could also be expected in the fully reduced samples.

\section{Discussion}
We state that bulk $R$NiO$_2$ samples show a universal spin glass behavior, without any sign of long-range magnetic orders\cite{LaNiO2, NdNiO2}. This is distinct from the parent cuprates which host AFM long-range order. 
The fitting parameters from the spin-dynamics models are similar for all $R$NiO$_2$ as well as Pr$_4$Ni$_3$O$_8$\cite{Pr438} (see Table \ref{tbl}), implying the similar spin-glass nature in these nickelates. According these results, the magnetic entities in $R$NiO$_2$ are expected to be cluster-like, rather than atomic single spins, and the interaction between the entities is not weak. 
Within the $R$NiO$_2$ system itself, we find that the relaxation times $\tau_0$ and $\tau_*$ become shorter and the energy barrier $E_a$ smaller from $R$=La, Pr to Nd. It is interesting to note that while the spin-freezing temperatures $T_f$ decrease from La, Pr to Nd, the critical temperatures in the Sr-doped thin-film versions of $R$NiO$_2$ increase (no superconductivity for $R$=La\cite{NdSrNiO2nature}, and $T_c$ = 7$\sim$12 K for $R$=Pr\cite{PrSrNiO2}, and $T_c$ = 9$\sim$15 K for $R$=Nd\cite{NdSrNiO2nature}. This has been attributed to a possible essential role of the 4$f$-electrons of the $R$ ions, and/or to stronger interactions between the Ni 3$d_{z^2}$ orbitals owing to a smaller ${c}$-axis lattice parameter\cite{Pickett}.

One question to be clarified is whether the observed spin-glass behaviour in the bulk $R$NiO$_2$ systems is intrinsic or not. First, the spin-glass behaviour is not likely due to elemental Ni particles, as is stated above. 
Second, we may refer to the strong similarity to the spin-glass dynamics in Pr$_4$Ni$_3$O$_8$\cite{Pr438} observed in very homogeneous single crystals. Given the fact that the electronic structure of the latter compound has been proposed to be very similar to that of NdNiO$_2$\cite{compare438and112}, we may conclude that the magnetic frustration causing the universal spin-glass behaviour in bulk $R$NiO$_2$ and perhaps even in Pr$_4$Ni$_3$O$_8$ has a common cause, and is not due to weak crystallization. Nevertheless, the fact that $T_s$ in the latter compound is almost one order of magnitude larger than in bulk $R$NiO$_2$, may hint to certain differences. 

Considering that Pr$_4$Ni$_3$O$_8$ is prepared by the topotactic reduction method similar to the title compounds, we cannot entirely rule out that the spin-glass behaviour could result from local non-stoichiometries of oxygen caused by incomplete or inhomogeneous reduction. An off-stoichiometric oxygen content can alter the coordination environment, thereby lead to distributed high-valence-state Ni ions and consequently make a frustration between various magnetic interactions possible. However, taking the essential role played by the interface and strain effect in nickelate thin films\cite{thickness} into account, the superconductivity along with the potential long-range magnetic order in thin films is not necessarily in contradiction to the occurrence of a disordered magnetism in bulk samples. With this respect, the spin-glass behaviour in bulk $R$NiO$_2$ can indeed be intrinsic. 
The spin glass state in $R$NiO$_2$ could arise from the self-doping effect of Nd 5$d$ electron pockets\cite{selfdope, EffectiveHamiltonian, hybridization}. 
Zhang $\textit{et al.}$\cite{selfdope},  proposed NdNiO$_{2}$ to be a self-doped Mott insulator, where the low-density Nd 5$d$ conduction electrons couple to the localized Ni 3$d_{x^2-y^2}$ electrons to form Kondo spin singlets at low temperatures. This self-doping effect suppresses the AFM long-range order. 
Choi $\textit{et al.}$\cite{flatband}, proposed that the Ni 3$d_{z^2}$ orbitals also participate strongly and forms a flat band pinned to the Fermi level in the nickelates. As a result, the magnetic orders are unstable and can be easily frustrated by the strong spin, charge, and lattice fluctuations, resulting in a spin disordered state but with AFM correlations\cite{Pickett}. Leonov $\textit{et al.}$\cite{Leonov}, also proposed that an unanticipated frustration of magnetic interactions in Nd$_{1-x}$Sr$_x$NiO$_2$ suppresses magnetic order.
Werner $\textit{et al.}$\cite{Werner}, argued that the optimally doped nickelate system is in a spin-freezing crossover region. Both overdoped and underdoped system are in a spin-freezing (or spin-glass-like) state, and the associated fluctuations of local moments may provide the pairing glue for the occurrence of superconductivity.

\section{Conclusions}
To conclude, we successfully synthesized bulk samples of the parent compounds $R$NiO$_2$ ($R$=La, Pr, Nd) and performed comprehensive magnetic measurements on them. The history-dependent DC magnetic susceptibilities, the magnetic hysteresis, the frequency dependence of the AC magnetizations, and more importantly, the memory effect reveal a universal spin-glass behaviour. According to phenomenological fitting results, we find that bulk $R$NiO$_2$ systems share a similarly slow spin dynamics as Pr$_4$Ni$_3$O$_8$, although with an almost one-order-of magnitude lower spin-freezing temperature. The magnetic entities are most probably cluster-like, and the interactions between these entities are not weak. Based on our investigations, we suggest that the universal spin glass behavior in bulk $R$NiO$_2$ may be due to off-stoichiometric oxygen or even to an intrinsic magnetic frustration, rather than to impurites or weak crystallization.

\section*{ACKNOWLEDGMENTS}
This work was supported by the Swiss National Foundation under Grants No. 20-175554, 206021-150784.

\end{spacing}

\pagebreak
\widetext
\title{Supplemental Material for\\
Universal spin-glass behaviour in bulk LaNiO$_{2}$, PrNiO$_{2}$ and NdNiO$_{2}$}\maketitle

\setcounter{equation}{0}
\setcounter{figure}{0}
\setcounter{table}{0}
\setcounter{page}{1}
\makeatletter
\renewcommand{\theequation}{S\arabic{equation}}
\renewcommand{\thefigure}{S\arabic{figure}}
\renewcommand{\bibnumfmt}[1]{[S#1]}
\renewcommand{\citenumfont}[1]{[S#1]}

\begin{figure}
  \includegraphics[width=12cm]{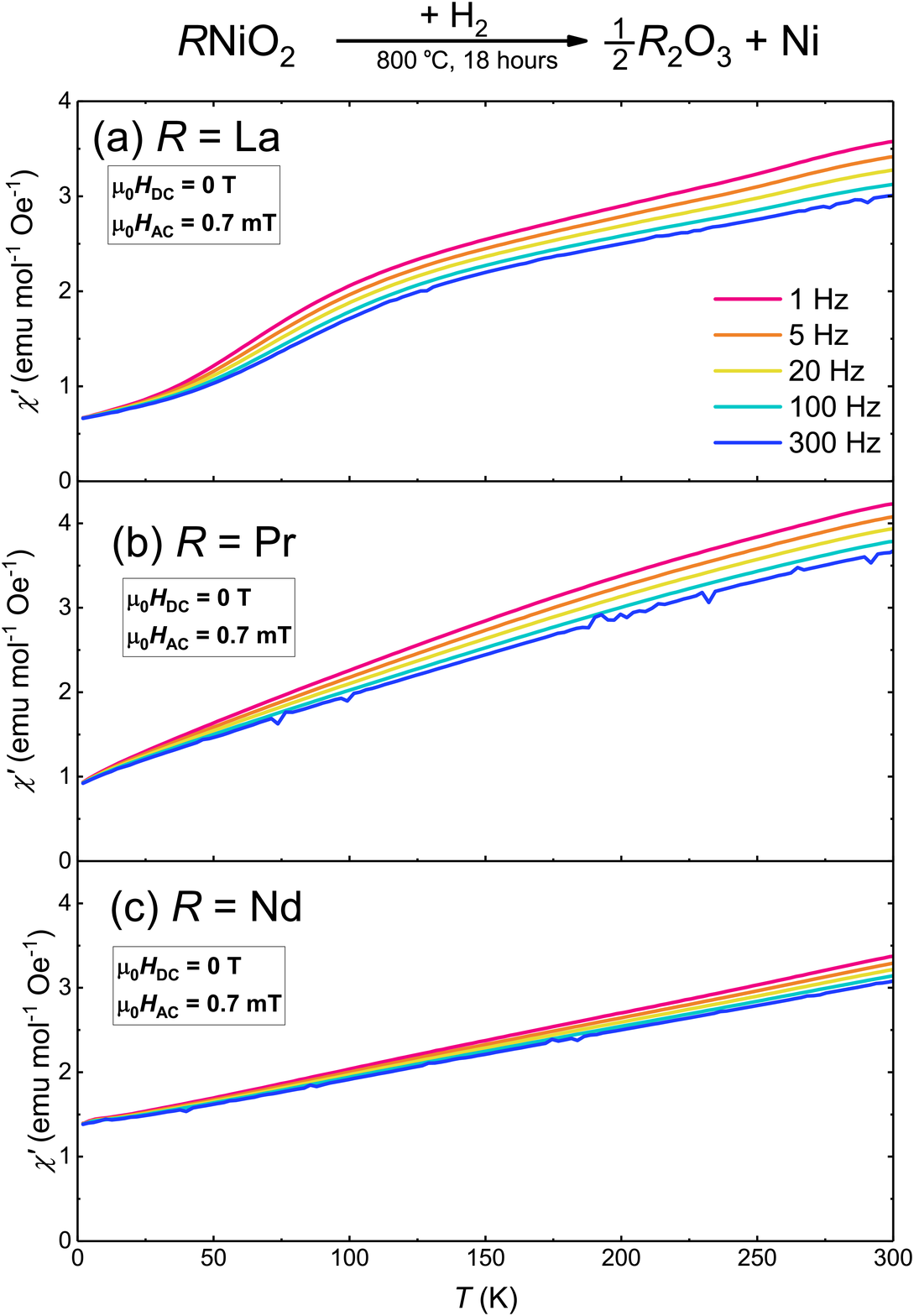}
\caption{AC magnetic susceptibilities measured on the fully-reduced products of bulk ${R}$NiO$_{2}$ samples, for ${R}$ = (a) La, (b) Pr, (c) Nd, respectively. Here, the products consist of the rare-earth oxides and the nickel. The data show the absence of spin-glass freezing peaks at low temperatures and the increasing frequency-dependence at high temperatures, both of which do not occur in the ${R}$NiO$_{2}$ samples. It supports that the spin-glass behaviours in the parent ${R}$NiO$_{2}$ do not originate from the impurities but from the intrinsic properties.} \label{figsi1}
\end{figure}

\end{document}